\documentclass[traditabstract]{aa}
\usepackage{txfonts}
\usepackage{graphicx}
\usepackage{amssymb}
\usepackage{url}
\usepackage[thicklines]{cancel}
\usepackage{natbib}
\bibpunct{(}{)}{;}{a}{}{,}
\newcommand{ \kms}{\mbox{km~s$^{-1}$}}
\newcommand{\mols}{\mbox{molec.~s$^{-1}$~}}
\renewcommand{\deg}{\mbox{$^{\circ}$}}
\begin{document}

%\thesaurus{10(03.07.1, 13.03.2, 18.05.1)}

\title{Ammonia and other parent molecules in comet 10P/Tempel 2 from
\textit{Herschel}/HIFI\thanks{{\em Herschel} is an ESA space observatory with
science instruments provided by European-led Principal Investigator consortia
and with important participation from NASA.} and ground-based radio
observations}

\author{N. Biver\inst{1} 
    \and J. Crovisier\inst{1}
    \and D. Bockel\'ee-Morvan\inst{1}
    \and S. Szutowicz\inst{2}
    \and D.C. Lis \inst{3} 
    \and P. Hartogh\inst{4}
    \and M. de Val-Borro\inst{4}
    \and R. Moreno\inst{1}
    \and J. Boissier\inst{5,6}
    \and M. Kidger\inst{7}
    \and M. K\"{u}ppers\inst{7}
    \and G. Paubert\inst{8}
    \and N. Dello Russo\inst{9}
    \and R. Vervack\inst{9}
    \and H. Weaver\inst{9}
    \and HssO team
}

\institute{LESIA, Observatoire de Paris, CNRS, UPMC, Universit\'e
Paris-Diderot, 5 place Jules Janssen, 92195 Meudon, France.
\email{Nicolas.Biver@obspm.fr}
\and Space Research Centre, PAS, Warszawa, Poland
\and Caltech, Pasadena, CA, USA
\and Max Planck Institut f\"ur Sonnensystemforschung, Katlenburg-Lindau, Germany
\and Istituto di Radioastronomia - INAF, Bologna, Italy
\and ESO,  Garching bei M\"unchen, Germany
\and ESAC, Villafranca del Castillo, Spain
\and IRAM, Avd. Divina Pastora, 7, 18012 Granada, Spain
\and JHU/APL, Laurel, Maryland, USA
}

% \date{Received  / Accepted}

\date{Draft \today}

\titlerunning{Ammonia in comet 10P/Tempel 2 with \textit{Herschel}}

\authorrunning{HssO team}

\abstract {
The Jupiter-family comet 10P/Tempel~2 was observed during its 2010 return with 
the \textit{Herschel Space Observatory}. We present here the observation of the
$J_{K}$ ($1_{0}$--$0_{0}$) transition of NH$_3$ at 572~GHz in this comet with the
Heterodyne Instrument for the Far Infrared (HIFI) of \textit{Herschel}. 
We also report on radio observations of other molecules (HCN, CH$_3$OH, 
H$_2$S and CS) obtained during the 1999 return of the comet with the CSO 
telescope and the JCMT, and during its 2010
return with the IRAM 30-m telescope. Molecular abundances relative to water are
0.09\%, 1.8\%, 0.4\%, and 0.08\% for HCN, CH$_3$OH, H$_2$S, and CS, 
respectively. 
An abundance of 0.5\% for NH$_3$ is obtained, which is similar to the values
measured in other comets. The hyperfine structure of the ammonia line is
resolved for the first time in an astronomical source. 
Strong anisotropy in the outgassing is present in all observations
from 1999 to 2010 and is modelled to derive the production rates. }

\maketitle

%\textbf{[J'ai chang\'e le titre et l'abstract. J'ai mis en gras les changements
%importants du texte (en plus des corrections d'orthographe et de typographie).]}

\section{Introduction}
%%%%%%%%%%%%%%%%%%%%%%

With an abundance of $\approx 0.5$\% relative to water, ammonia is a
major repository of nitrogen in cometary volatiles \citep{Boc04}. The
photodissociation products of NH$_{3}$, NH$_{2}$ and NH, are routinely observed
in the visible spectra of comets \citep{Fel04}. The direct observation of 
ammonia is difficult because it is a short-lived molecule (lifetime 
$\approx 5000$~s at 1 AU from the Sun) and because its lines are either 
weak or affected by telluric absorption.

In previous radio observations of NH$_3$, the inversion transitions near 
24~GHz were tentatively detected with ground-based radio telescopes in C/1983 H1
(IRAS-Araki-Alcock) \citep{Alt83}, but not in 1P/Halley \citep{Bir87}.
They were then definitely detected in C/1996 B2 (Hyakutake) \citep{Pal96}
and C/1995 O1 (Hale-Bopp) \citep{Bir97, Hir99} and tentatively
detected in 153P/Ikeya-Zhang \citep{Bir02, Hat05}. 

Ammonia rotational lines are expected to be much stronger, but they fall in the
submillimetric spectral range and have to be observed from space. The $J_{K}$
($1_{0}$--$0_{0}$) line at 572.5~GHz was first detected in C/2001 Q4 (NEAT)
and C/2002 T7 (LINEAR) with the Odin satellite \citep{Biv07}.

Ammonia was also observed from its $\nu_{1}$ and/or $\nu_{3}$ vibrational bands
near 3~$\mu$m in comets 6P/d'Arrest, 73P/Schwassmann-Wachmann 3, C/1995
O1 (Hale-Bopp), C/2002 T7 (LINEAR), C/2004 Q2 (Machholz), C/2006 P1 (McNaught),
and 103P/Hartley~2 \citep[ and references therein]{Kaw11}.

Comet 10P/Tempel~2 is a Jupiter-family comet discovered in 1873 by
Wilhelm Tempel. Its orbital period is 5.5 years; thus alternate
perihelion passages are favourable. Its behaviour has already been well
documented, 2010 being the year of the 22$^{\rm nd}$ return to perihelion 
observed for this comet. Prior to 1994, its perihelion distance was shorter 
($\approx$~1.31--1.39 AU vs 1.48 AU), resulting in a closer approach to the 
Earth and the comet being more active at perihelion. After the 1999 apparition,
the orbit of the comet slightly changed and the perihelion distance
decreased again to 1.42 AU.

Comet 10P/Tempel~2 has a relatively large nucleus \citep[$\approx16\times8$
km;][]{Jew89,Lam09}, and a well-studied rotation period of
$\approx$~8.95~h, found to be increasing with time \citep{Kni11}. Given its
relatively low outgassing rate, only a small fraction of its nucleus
is active. Strong seasonal effects are observed: the activity rises very 
rapidly during the last three months before perihelion and peaks about 
one month after. A prominent
dust jet close to the north pole has been observed in images at each perihelion.

We report on an observation of the 572.5~GHz line of ammonia 
in comet 10P/Tempel~2, which was part of the study of this comet with the
\textit{Herschel Space Observatory}. Preliminary reports were given by
\citet{Biv10} and \citet{Szu11}. 
Additional analyses of the observations of water with 
the Heterodyne Instrument for the Far Infrared (HIFI) in this comet 
will be presented by \citet{Szu12}. 
Several water lines were also detected with the 
Photodetector Array Camera and Spectrometer (PACS) and 
the Spectral and Photometric Imaging Receiver (SPIRE) instruments
of Herschel and will be presented in a future paper.

In support of Herschel observations, 10P/Tempel~2 was observed from the
ground at millimetric wavelengths with the Institut de radioastronomie
millim\'etrique (IRAM) 30-m telescope. We also present observations obtained 
at the antepenultimate perihelion passage (in 1999), 
with the Caltech Sumillimeter Observatory (CSO) telescope and the 
James Clerk Maxwell Telescope (JCMT), which helped to prepare the Herschel 
observations.

\section{Observations}
%%%%%%%%%%%%%%%%%%%%%%

\begin{figure}[h]
%\centerline{\includegraphics[width=\columnwidth, angle=270]{10phcn32a}}
\resizebox{\hsize}{!}{\includegraphics[angle=270]{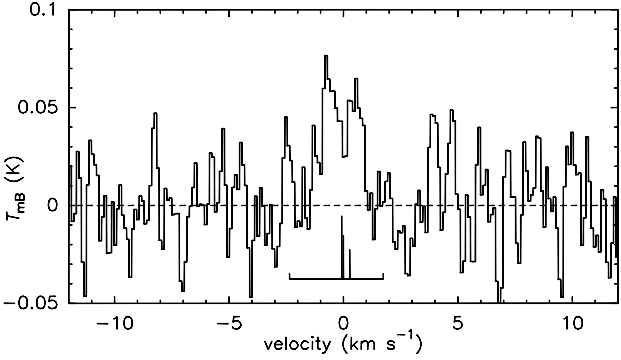}}
\caption{HCN $J$(3--2) line at 265.886 GHz observed in comet
10P/Tempel~2 with the CSO on 11--12 September 1999.
The positions and relative intensities of the hyperfine components are drawn
as detailed in Fig.~\ref{10phcn32b}.}
\label{10phcn32a}
\end{figure}

The 1999 perihelion of comet 10P/Tempel~2 was on 8 September at $r_h = 1.48$~AU.
The perigee took place on 12 July at 0.65~AU. 
HCN was observed at JCMT on 4 and 6 September, and again on 1 and 2 October. 
HCN was also observed at CSO on 11 and 12 September (Fig.~\ref{10phcn32a}) and 
CH$_3$OH on the 12th only. 

\begin{figure}[h]
\resizebox{\hsize}{!}{\includegraphics[angle=270]{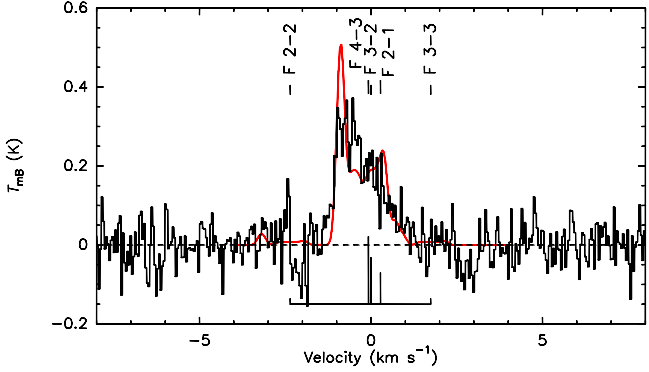}}
\caption{HCN $J$(3--2) line at 265.886 GHz observed in comet
10P/Tempel~2 with the IRAM 30-m telescope on 10--11 July 2010.
The position and relative intensities of the hyperfine components are shown
and a synthetic spectrum including asymmetric outgassing (see text) and
hyperfine structure is overplotted.}
\label{10phcn32b}
\end{figure}

\begin{figure}[h]
\resizebox{\hsize}{!}{\includegraphics[angle=270]{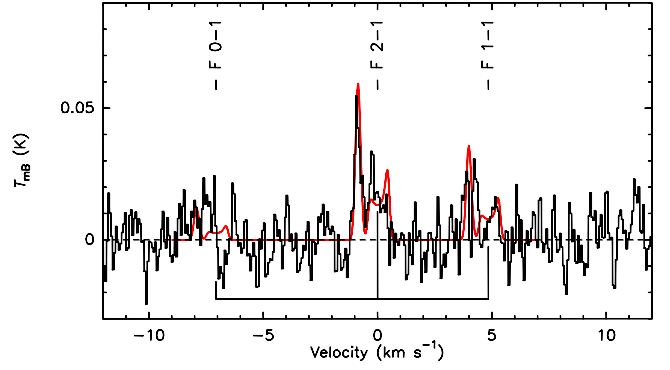}}
\caption{HCN $J$(1--0) line at 88.632 GHz observed in comet
10P/Tempel~2 with the IRAM 30-m telescope on 7--11 July 2010.
The position and relative intensities of the hyperfine components are shown
and a synthetic spectrum considering asymmetric outgassing (see text) 
is overplotted.}
\label{10phcn10}
\end{figure}

\begin{figure}[h]
\resizebox{\hsize}{!}{\includegraphics[angle=270]{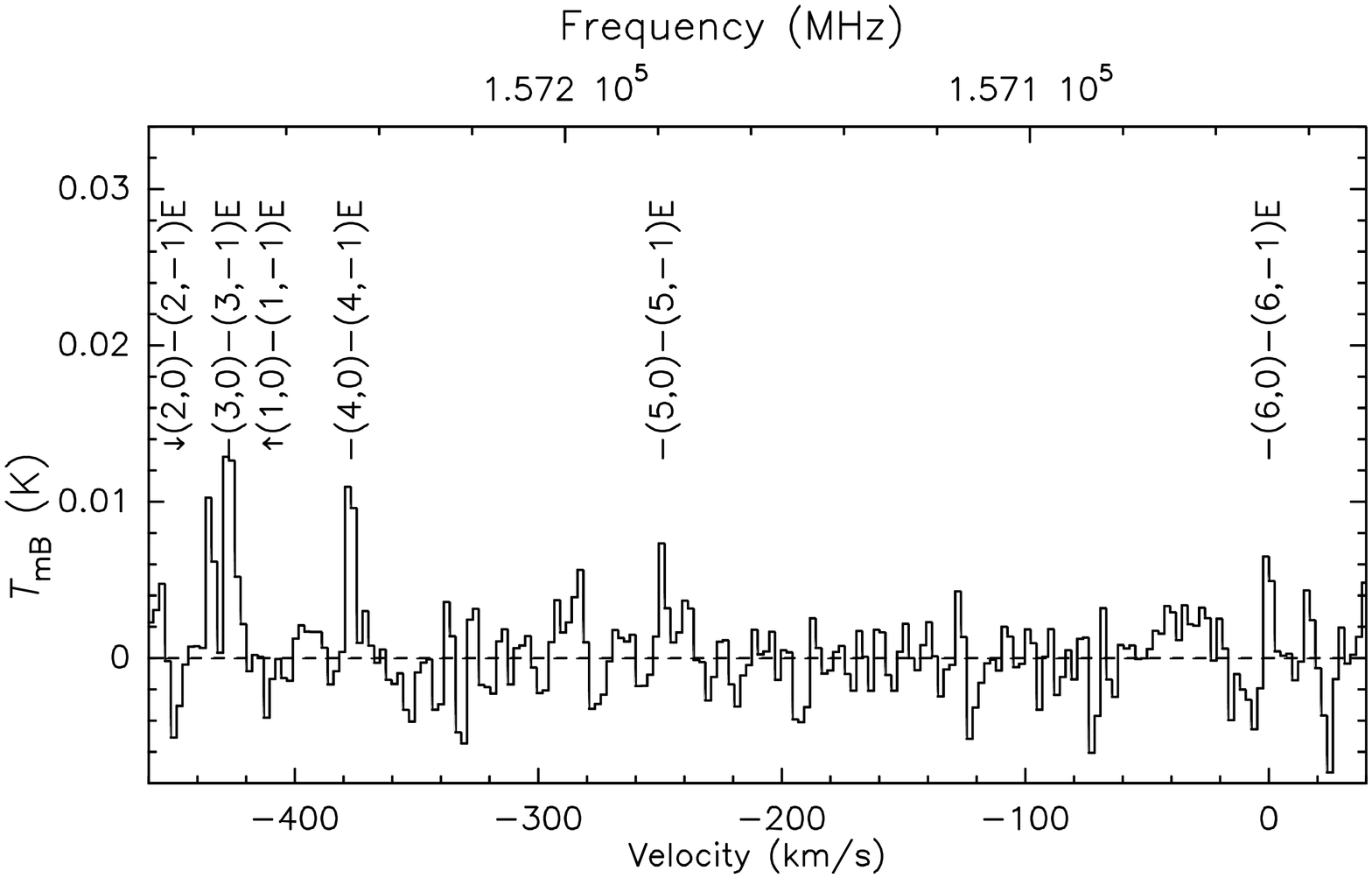}}
\caption{CH$_3$OH lines at 157~GHz observed in comet
10P/Tempel~2 with the IRAM 30-m telescope on 11 July 2010.}
\label{10pch3oh}
\end{figure}

The 2010 perihelion was on 4 July at $r_h = 1.42$~AU and perigee took place
on 26 August at $\Delta = 0.65$~AU. The comet was observed with the IRAM 30-m 
radio-telescope between 7.2 and 11.4 July 2010 UT. The
first two nights suffered from bad weather and the third was completely lost, 
but days 4 and 5 yielded good data (Figs.~\ref{10phcn32b}--\ref{10pch3oh}).
Table~\ref{tabobs} provides the list of detected lines.

\begin{figure}[h]
\resizebox{\hsize}{!}{\includegraphics[angle=270]{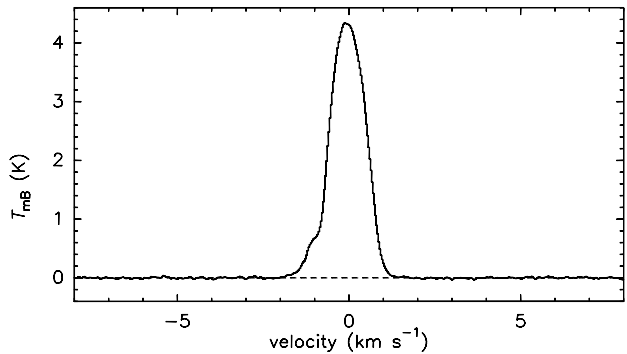}}
\caption{Water H$_2$O($1_{10}$--$1_{01}$) line observed in comet
10P/Tempel~2 with \textit{Herschel} using the HRS on 19 July 2010.}
\label{10ph2o}
\end{figure}

\begin{figure}[h]
\resizebox{\hsize}{!}{\includegraphics[angle=270]{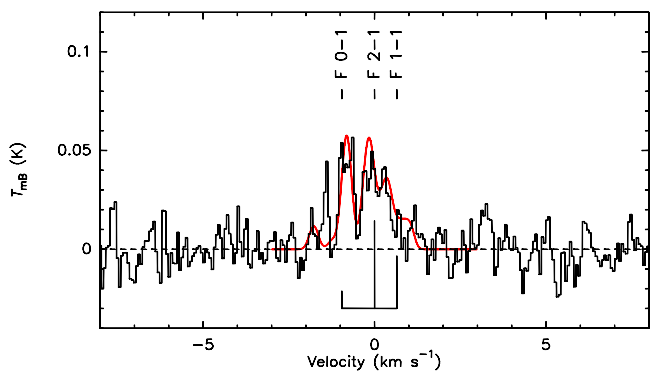}}
\caption{Ammonia $J_{K}$ ($1_{0}$--$0_{0}$) line observed in comet
10P/Tempel~2 with \textit{Herschel} using the HRS on 19 July 2010.
The position and relative intensities of the hyperfine components are shown
with a synthetic spectrum overplotted that takes into account hyperfine 
structure and asymmetric outgassing (see text).}
\label{10pnh3}
\end{figure}

Comet 10P was also observed in 2010 with the \textit{Herschel Space Observatory}
\citep{Pil10} using HIFI
\citep{deGra10}, within the framework of the \textit{Water and related 
chemistry in the Solar System} (HssO) project \citep{Har09}. 
Several water rotational lines were observed and mapped from 15 June to 29
July 2010 \citep{Szu11,Szu12}. 
The observation 
of ammonia took place on 19.1 July 2010 UT when the comet was at 
$r_{h} =1.43$~AU and at $\Delta = 0.71$~AU from Herschel.
The $J_{K} $($1_{0}$--$0_{0}$) transition of NH$_3$ at 572.5~GHz and the 
$1_{10}$--$1_{01}$ line of H$_2$O at 557~GHz were observed simultaneously
in the frequency-switching mode, the former in the upper side band and the 
latter in the lower side band of the band 1b receiver of HIFI. 
A total of 54 min of integration time were spent on the
comet, split into five observations every 2~h to cover a full rotation of the
nucleus. Water and ammonia were observed both with the low-resolution 
spectrometer (WBS, resolution 0.58~\kms) and the highest resolution mode of 
the high-resolution spectrometer (HRS autocorrelator, resolution 0.074~\kms) 
and both polarizations were averaged (Figs~\ref{10ph2o},~\ref{10pnh3}).

\begin{table*}
\caption[]{Molecular observations in comet 10P/Tempel 2} \label{tabobs}
\begin{tabular}{lcclrclcc}
\hline
\hline
  & UT date & $<r_{h}>$ & $<\Delta>$ & Integ. time & Line & Intensity   & Velocity shift$^b$ &  Prod. rate\\[0.cm]
  & [mm/dd.dd] & [AU]  & [AU]       & [min]$^a$ &     &    [K~ \kms]    &   [ \kms]   &  [\mols] \\
\hline \\[0.1cm]
\multicolumn{8}{l}{1999 perihelion passage:} \\
\hline
JCMT & 09/04.2--06.3 & 1.482 & 0.816 & 157 & HCN(4--3) & $0.058\pm0.012$ & $-0.86\pm0.23$ & $0.5\pm0.1\times10^{25}$\\
CSO  & 09/11.2--12.3 & 1.482 & 0.851 & 133 & HCN(3--2) & $0.106\pm0.012$ & $-0.20\pm0.08$ & $1.4\pm0.2\times10^{25}$\\
CSO  & 09/12.32      & 1.482 & 0.855 &  72 & CH$_3$OH($4_1-4_0$A-+) & $0.080\pm0.018$ & $+0.19\pm0.15$ & \multicolumn{1}{|l}{} \\
     &               &    & & & CH$_3$OH($2_1-2_0$A-+) & $0.119\pm0.016$ & $-0.10\pm0.08$ & \multicolumn{1}{|c}{$3.7\pm1.0\times10^{26}$}\\
JCMT & 10/01.3--02.3 & 1.500 & 0.980 &  67 & HCN(3--2) & $0.090\pm0.019$ & $-0.57\pm0.18$ & $0.9\pm0.2\times10^{25}$\\
\hline \\[0.1cm]
\multicolumn{8}{l}{2010 perihelion passage:} \\
\hline
 IRAM & 07/07.28      & 1.423 & 0.745 & 126 & HCN(1--0)   & $0.057\pm0.013$ & $-0.16\pm0.16$ & $1.6\pm0.4\times10^{25}$\\
 IRAM & 07/08.26      & 1.423 & 0.742 & 131 & HCN(1--0)   & $0.054\pm0.009$ & $-0.44\pm0.14$ & $1.5\pm0.2\times10^{25}$\\
 IRAM & 07/10.22      & 1.424 & 0.734 & 103 & HCN(3--2)   & $0.410\pm0.033$ & $-0.30\pm0.05$ & $1.5\pm0.1\times10^{25}$\\
 IRAM & 07/10.34      & 1.424 & 0.734 &  42 & HCN(1--0)   & $0.044\pm0.016$ & $-0.65\pm0.33$ & $1.2\pm0.4\times10^{25}$\\
 IRAM & 07/11.17      & 1.424 & 0.731 &  61 & HCN(3--2)   & $0.427\pm0.036$ & $-0.41\pm0.05$ & $1.6\pm0.2\times10^{25}$\\
 IRAM & 07/11.28      & 1.424 & 0.731 & 131 & HCN(1--0)   & $0.068\pm0.010$ & $-0.49\pm0.13$ & $1.8\pm0.3\times10^{25}$\\
 IRAM & 07/07.2-08.4  & 1.423 & 0.743 & 257 & CH$_3$OH($1_0-1_{-1}$E) &  $0.028\pm0.010$ & \multicolumn{1}{|c|}{} &  \\
      &               &       &       & & CH$_3$OH($2_0-2_{-1}$E) &  $0.044\pm0.010$ & \multicolumn{1}{|c|}{} &  \\
      &               &       &       & & CH$_3$OH($3_0-3_{-1}$E) &  $0.032\pm0.010$ & \multicolumn{1}{|c|}{} &  \\
  & & & & & CH$_3$OH($4_0-4_{-1}$E) &  $0.050\pm0.010$ & \multicolumn{1}{|c|}{$-0.41\pm0.10^c$} &  $3.2\pm0.8\times10^{26}$\\
      &               &       &       & & CH$_3$OH($5_0-5_{-1}$E) &  $0.030\pm0.010$ & \multicolumn{1}{|c|}{} &  \\
      &               &       &       & & CH$_3$OH($6_0-6_{-1}$E) &  $0.036\pm0.010$ & \multicolumn{1}{|c|}{} &  \\
      &               &       &       & & CH$_3$OH($7_0-7_{-1}$E) &  $0.008\pm0.010$ & \multicolumn{1}{|c|}{} &  \\
&&&&&&&\\[-0.3cm]
 IRAM & 07/11.26      & 1.424 & 0.731 &  98 & CH$_3$OH($1_0-1_{-1}$E) &  $0.024\pm0.008$ & \multicolumn{1}{|c|}{} &  \\
      &               &       &       &  & CH$_3$OH($2_0-2_{-1}$E) &  $0.046\pm0.008$ & \multicolumn{1}{|c|}{} &  \\
      &               &       &       &  & CH$_3$OH($3_0-3_{-1}$E) &  $0.050\pm0.008$ & \multicolumn{1}{|c|}{} &  \\
  & & & & & CH$_3$OH($4_0-4_{-1}$E) &  $0.041\pm0.008$ & \multicolumn{1}{|c|}{$-0.42\pm0.08^c$} &  $3.2\pm0.4\times10^{26}$\\
      &               &       &       &  & CH$_3$OH($5_0-5_{-1}$E) &  $0.030\pm0.008$ & \multicolumn{1}{|c|}{} &  \\
      &               &       &       &  & CH$_3$OH($6_0-6_{-1}$E) &  $0.017\pm0.008$ & \multicolumn{1}{|c|}{} &  \\
      &               &       &       &  & CH$_3$OH($7_0-7_{-1}$E) &  $0.011\pm0.008$ & \multicolumn{1}{|c|}{} &  \\
 IRAM & 07/10.34      & 1.424 & 0.734 &  42 & H$_2$S($1_{10}-1_{01}$) & $0.065\pm0.021$ & $-0.43\pm0.26$ & $7.1\pm2.3\times10^{26}$\\
 IRAM & 07/11.34      & 1.424 & 0.730 &  33 & CS(3--2)    & $0.049\pm0.011$ & $-0.02\pm0.14$ & $1.5\pm0.3\times10^{25}$\\
 IRAM & 07/11.34      & 1.424 & 0.730 &  33 & CH$_3$CN(8$_K$-7$_K$) $K$=0,1,2$^d$    & $<0.062$ &     & $<0.8\times10^{25}$\\
\hline
 HIFI & 07/18.9--19.3 & 1.431 & 0.708 &  54 & H$_2$O($1_{10}$--$1_{01}$) & $5.509\pm0.005$ & $-0.062\pm0.001$ & $2.2\pm0.1\times10^{28}$\\
 HIFI & 07/18.9--19.3 & 1.431 & 0.708 &  54 & NH$_3$($1_0$--$0_0$) & $0.072\pm0.005$ & $-0.43\pm0.06^e$ & $10.0\pm0.7\times10^{25}$\\
\hline
\end{tabular}
\\
Note: $^a$ Total integration time, ON+OFF, or ON in frequency switching mode (JCMT and HIFI); \\
$^b$ Reference line frequencies (excepted for NH$_3$ -- see text$^e$) were taken
from CDMS \citep{CDMS}; \\
$^c$ value based on the average of the 6 strongest lines; \\
$^d$ Sum of the three transitions $J_K=$ 8$_0$--7$_0$, 8$_1$--7$_1$, and 8$_2$--7$_2$; \\
$^e$ Doppler shift measured with respect to the barycentric position of
the hyperfine structure of the line at 572498.160~MHz.
\end{table*}

\section{Results}
%%%%%%%%%%%%%%%%%

\subsection{Modelling the data}
\label{section:model}

The same excitation and outgassing pattern model has been used 
to analyse all data. The density distribution is based on the Haser model, 
with photodissociation rates at 1 AU, $\beta_0\approx1.88\times10^{-5}$
s$^{-1}$ in 1999 and $\beta_0=1.60\times10^{-5}$ s$^{-1}$ in 2010,
for HCN, based on the solar activity. The gas temperature is constrained
by the rotational temperature of methanol lines and the outgassing
pattern and expansion velocity are constrained by the line shapes.
Collisions with electrons are taken into account as modelled in 
\citet{Zak07} with a density scaling factor $x_{ne}$ of 0.5 \citep{Biv06}.  

All lines (Figs.~\ref{10phcn32a}--\ref{10phcn10})
show a strong asymmetry (mean Doppler shift around $-0.3$ to $-0.4$ \kms,
Table~\ref{tabobs}) and need to be modelled with asymmetric outgassing.
The strong blueshift and a phase angle ($\approx$42\deg) smaller than 90\deg~
suggest that a large part of the outgassing is dominated by a relatively
narrow jet on the day-side that is not far from the direction to the Earth.
The opening of the jet and the fraction of outgassing it contains were
constrained by the Doppler shifts of the lines and optimized to fit the line 
shapes. The half width at half maximum (HWHM) of the lines on the negative 
and positive sides suggests expansion velocities of 0.9 and 0.5~\kms~ 
on the day and night sides, respectively.

The rotational temperature of methanol was found to be $16\pm7$~K, $32\pm8$,
and $25\pm4$~K on 7--8 September 1999, and 11 July 2010 
(Fig.~\ref{10pch3oh}), respectively. 
These values are compatible with a gas temperature
$T_\mathrm{rot} = 25$~K, but a closer inspection of the relative intensities
of the lines reveals that the night-side temperature 
($v > 0$~\kms) $T_\mathrm{rot}=21\pm5$~K is lower than in the
day-side jet: $T_\mathrm{rot}=31\pm4$~K. We will use $T_\mathrm{rot}=30$~K in
the sunward jet and $T_\mathrm{rot}=20$~K elsewhere. 

\begin{figure}[h]
\resizebox{\hsize}{!}{\includegraphics[angle=270]{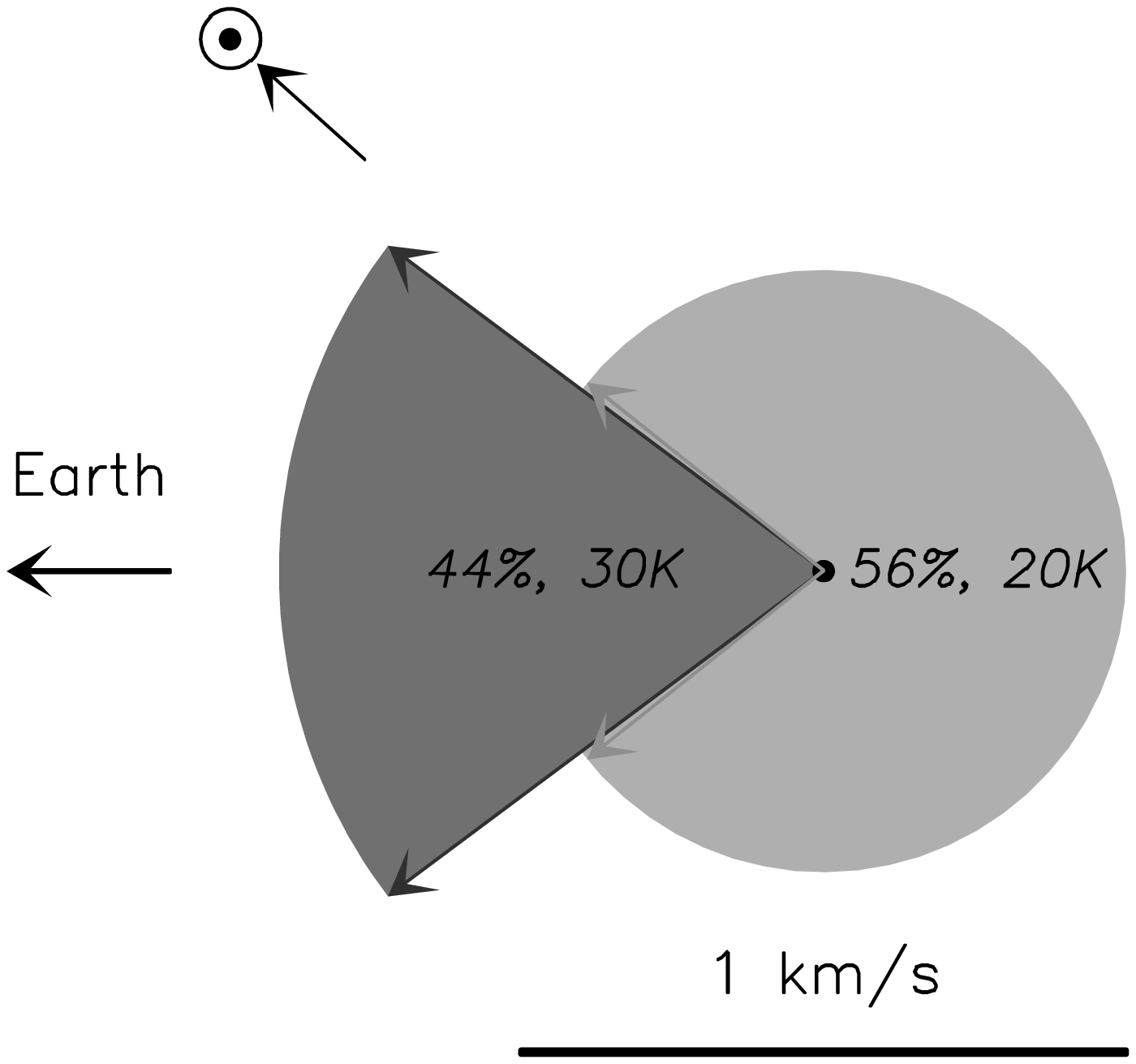}}
\caption{Outgassing pattern
used to interpret 10P/Tempel~2 observations. The length of the 
vectors pointing away from the nucleus are proportional to the 
expansion velocity (scale bar).
The shaded area in dark
represents the earthward jet at 0.9~\kms containing 44\% of the outgassing with
a $T=30$ K temperature, while the lightly shaded area represents the
remaining part of the sky with lower velocity (0.5~\kms) and temperature
(20 K).}
\label{10pjet}
\end{figure}

A good fit is obtained with 44\% of the outgassing in a narrow jet 
(opening angle 37\deg~ or $0.1\times4\pi$ steradian) with gas velocity of 
0.9~\kms, and the remaining 90\% of the sky contain 56\% of
the outgassing at 0.5~\kms, as illustrated in Fig.~\ref{10pjet}. 
The synthetic line shapes obtained with this
modelling provide a reasonable fit to the observed lines
(Figs.~\ref{10pnh3},~\ref{10phcn32b} and \ref{10phcn10}). 
We used the same model to analyse the H$_2$O 557~GHz line
(Fig.~\ref{10ph2o}) and determine the water production rate 
$Q_{\rm H_2O}$ at the time
of the NH$_3$ observations. However, we assumed $x_{ne} = 0.15$ in the
excitation model. This value best explains the brightness distribution
of the 557~GHz line observed on 19 July \citep{Szu11,Szu12} and is consistent
with values found in other comets \citep{Biv07,Har10}.
The model reproduces both the line intensity and the Doppler
shift of the water line if we set $Q_{\rm H_2O}$ to $2.2\times10^{28}$~\mols 
(Table~\ref{tabobs}). 
This simplistic modelling of the
anisotropic outgassing of comet 10P/Tempel~2 will be improved in a future
paper \citep{Szu12}, but is sufficient to derive accurate 
production rates and relative abundances. When isotropic 
outgassing is assumed and the gas temperature set to 25~K,
we derive production rates $\approx$15\% higher 
for the short-lived molecules like H$_2$S and NH$_3$, 
or $\approx$45\% higher for other molecules.

\subsection{Ammonia} 

The ammonia $J_{K}$ ($1_{0}$--$0_{0}$) line at 572.5~GHz is clearly detected
(Fig.~\ref{10pnh3}). The hyperfine structure of the ammonia line is
similar to that of the $J$~(1--0) HCN line: three components $F$~(2--1),
$F$~(1--1) and $F$~(0--1) with relative intensities 5, 3 and 1, and relative
velocities at 0.0, $+0.64$ and $-0.97$~km~s$^{-1}$, respectively,
with respect to the $F$~(2--1) frequency of 572498.371~MHz \citep{Caz09}.
This structure is resolved in the observed spectrum of NH$_3$
(Fig.~\ref{10pnh3}). The synthetic line profile, based on a model used to fit
the optically thin molecular lines observed at IRAM (Figs~\ref{10phcn32b} and
~\ref{10phcn10}, Section~\ref{section:model}), 
agrees well with the observed profile.

\begin{figure}[h]
\resizebox{\hsize}{!}{\includegraphics[angle=270]{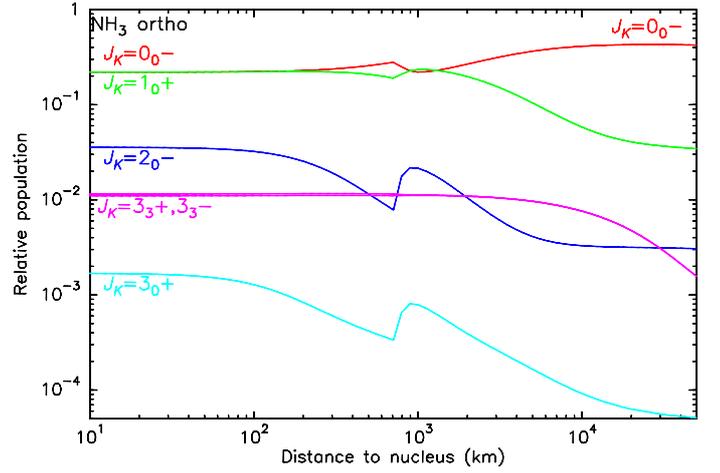}}
\caption{Rotational populations of the lowest ortho levels of NH$_3$
in the coma of comet 10P/Tempel~2 
($r_h=1.431$ AU, $Q_{\rm H_2O}=2.2\times10^{28}$~\mols, $v_{exp}=0.7$~\kms), 
taking into account collisions with neutrals at $T=25$ K, 
collisions with electrons (responsible for the
feature at the contact surface at $\approx1000$ km), 
radiative decay and infrared pumping.}
\label{nh3pop}
\end{figure}

The time variation of the line intensity is marginal ($\pm17$\%,
at a 1.5--$\sigma$ level of significance). The ammonia production
rate was derived with the previously described collision and outgassing pattern
model.  We assumed a total cross-section of $2\times10^{-14}$~cm$^{-2}$ 
for the collisions with neutrals. Collisions with electrons 
were modelled in the same way as for the other molecules \citep[e.g.][]{Zak07},
i.e. we used the Born approximation
for the computation of the cross-sections and set $x_{ne}=0.5$. 
Infrared pumping through the six strongest ($\nu_1$, $\nu_2$, $\nu_3$,
$\nu_4$, $\nu_3+\nu_4$, and $\nu_2+\nu_3$) vibrational bands was
taken into account using the GEISA database \citep{Geisa}. 
Table~\ref{tabnh3} lists the pumping rates for the main infrared
bands of ammonia. The six bands included in our model comprise 95\% of the 
infrared pumping. Our excitation rates agree with those of 
\citet{Kaw11}. When computing the fluorescence of the vibrational bands 
down to the fundamental ground state, we did not consider that the
$\nu_3+\nu_4$, and $\nu_2+\nu_3$ bands partly deexcite via $\nu_3$.
Since these combination bands are weak, this approximation should not 
affect significantly the rotational populations.
The radial evolution of the population of the six
lowest ortho levels of NH$_3$ is shown in Fig.~\ref{nh3pop}.

The NH$_3$ photodissociation rate is $\beta_0 = 1.8 \times 10^{-4}$~s$^{-1}$ 
for the quiet Sun at $r_{h} = 1$~AU \citep{Hue92}. 
Because the NH$_3$ lifetime is relatively short, radiative processes
do not significantly affect the excitation of the rotational levels for comets
with high outgassing rates. Indeed NH$_3$ photodissociates before reaching
the rarefied coma where IR pumping and radiative decay dominate over 
collisional excitation. For moderately active comets like 
10P/Tempel~2, these processes are significant: 
neglecting IR pumping would overestimate the
production rate by a factor 2.5 and assuming thermal equilibrium would
underestimate it by a factor 1.7. On the other hand, neglecting the 
anisotropy of the outgassing would increase the production rate by only 
14\%. The derived ammonia production rate is
$1.0\times10^{26}$ \mols. The contemporaneous water
production being $\approx 2.2\times10^{28}$ \mols (Table~\ref{tabobs}),
we obtain [NH$_{3}$]/[H$_{2}$O] = $0.46\pm0.04$\%. This value is
very similar to the abundance measured in other comets 
\citep{Kaw02, Biv07, Kaw11}.

The ortho-to-para ratio (OPR) of ammonia is not directly measured in comets; it
is inferred from that of the NH$_2$ radical and found to be typically 1.1 to
1.2, corresponding to spin temperatures $\approx 30$~K \citep{Kaw01, Shi10,
Shi11}. When analysing our measurement, where only \textit{ortho} NH$_3$ was
observed, we assumed OPR = 1 (statistical ratio). Therefore, our determination
of the production rate may be overestimated by 5 to 10\%.

\begin{table}
\caption[]{Ammonia infrared band parameters and excitation rates} \label{tabnh3}
\begin{tabular}{lccrc}
\hline
\hline
vibrational band & frequency  & $A_{\nu}^a$ & $g_{\nu}^a$ at 1 AU & used \\[0.cm]
                 & $[$cm$^{-1}]$ & $[$s$^{-1}]$ & $[$s$^{-1}]$ &  \\
\hline
$\nu_1$       &          3337 &   7.7   &  $3.3\times10^{-5}$ & yes \\
$\nu_2$       &  $\approx$950 &  16.2   & $27.7\times10^{-5}$ & yes \\
$2\nu_2$      & $\approx$1700 &  0.5   &  $0.5\times10^{-5}$ & no \\
$3\nu_2$      & $\approx$2600 &  0.1   &  $0.05\times10^{-5}$ & no \\
$\nu_1+\nu_2$ &          4315 &  1.6   &  $0.5\times10^{-5}$ & no \\
$\nu_3$       &          3450 &   3.8   &  $1.5\times10^{-5}$ & yes \\
$\nu_4$       &          1630 &   8.7   &  $8.5\times10^{-5}$ & yes \\
$\nu_2+\nu_4$ &    2600 &   0.02   &  $0.01\times10^{-5}$ & no \\
$\nu_3+\nu_4$ &    5000 &   15    &  $3.7\times10^{-5}$ & yes \\
$\nu_1+\nu_4$ &    4960 &   0.6   &  $0.2\times10^{-5}$ & no \\
$\nu_2+\nu_3$ &    4450 &    9    &  $2.6\times10^{-5}$ & yes \\
$2\nu_4$      &    3240 &   2.7   &  $0.4\times10^{-5}$ & no \\
%$\nu_1+\nu_3$  & 1,0,1,0  &    6000 &   0   &  $0\times10^{-5}$ & no \\
\hline
\end{tabular}
\\
Note: $^a$ ``Band'' \citep[as defined in][]{Cro83} Einstein coefficients 
and total excitation rates (from GEISA database \citep{Geisa}) computed at 40 K.
\end{table}

\subsection{Molecular abundances from ground based data}

Table~\ref{tabobs} provides production rates inferred for
HCN and CH$_3$OH in 1999, and for HCN, CH$_3$OH, CS, H$_2$S and
 CH$_3$CN in 2010. Based on contemporaneous water production 
of $\approx 1.8\times10^{28}$~\mols \citep{Szu12} at the time
of IRAM observations in 2010 (i.e. $\approx$ 9 days earlier than the 
NH$_3$ observation), the derived molecular abundances are given in
Table~\ref{tababund}. These abundances are typical of short-period comets, with
CH$_3$OH and H$_2$S abundances in the middle-low part of the observed range
\citep{Cro09a,Cro09b}. The average cometary [CS]/[HCN] ratio is close 
to 0.8 at $r_h = 1$ AU, with a $r_h^{-0.8}$ dependence \citep{Biv06,Biv11}.
The observed value in 10P/Tempel~2 is on the high side given
that we would expect a value around 0.6 at $r_h = 1.42$ AU.
Ammonia is the dominant source of nitrogen in the cometary
ices ($\approx$ 80\%). 
Other possible sources of nitrogen in comets are HNC,
and HC$_3$N, which were always below the HCN abundance
in all comets in which they were detected or searched for
\citep[e.g.,][]{Boc04}.
Note that even though HC$_3$N was not searched in depth in comet
10P, low-resolution spectra on 11.3 July provide an upper limit
on the HC$_3$N $J$(16--15) line, which yields in any 
case $Q_{\rm HC_3N} < 0.3\times Q_{\rm NH_3}$.

\begin{table}
\caption[]{Molecular abundances in comet 10P/Tempel 2} \label{tababund}
\begin{tabular}{lccc}
\hline
\hline
Molecule & $Q/Q_{\rm HCN}$ & $Q/Q_{\rm HCN}$ & $Q/Q_{\rm H_2O}$ \\[0.cm]
         &  in 1999   & in 2010 & in 2010 \\
\hline
HCN      &    1     &     1          &  $0.09\pm0.01$\% \\
CH$_3$OH & $20\pm5$ & $21\pm3$       &  $1.8\pm0.2$ \% \\
H$_2$S   &          & $4.6\pm1.6$    &  $0.39\pm0.13$ \% \\
CS       &          & $0.9\pm0.2$    &  $0.08\pm0.02$ \% \\
CH$_3$CN &          & $<0.5$         &  $<0.04$ \% \\
NH$_3$   &          &     -          &  $0.46\pm0.04$ \% \\
\hline
\end{tabular}
\end{table}

%\section{Discussion}
%%%%%%%%%%%%%%%%%%%%

In 1999 the comet had a higher activity on 12 September when CH$_3$OH
was observed and HCN reached a maximum. At other dates, however, it was at least
50\% less active than in 2010. This is likely due to a larger perihelion
distance (1.48 vs 1.42 AU). Meanwhile, the [CH$_3$OH]/[HCN] ratio did not change
in 11 years after two orbits around the Sun and the prominent jet
feature was still there: the line asymmetry is still present 
and visible images showed a similar northwards jet structure
(Biver personal communication) at both apparitions.

\section{Conclusion}
%%%%%%%%%%%%%%%%%%%%

Ammonia was directly observed in comet 10/Tempel~2 through its fundamental
rotational line. The hyperfine structure of the line is resolved for the first
time in an astronomical object, thanks to the narrow width of the cometary
line. The abundance of ammonia is found to be $0.46 \pm 0.04$\% relative
to water. This is similar to values measured in other comets, making
ammonia the major source of nitrogen in cometary ices.
The abundances of HCN, CH$_3$OH, CS, and H$_2$S are typical
of short-period comets, though on the mid-low range.
The comet displayed strongly blueshifted lines indicative of a
strong asymmetry in the outgassing. This asymmetry is present for all species,
which also suggests a common source for all and compositional homogeneity. 
Based on simple modelling, about half of the outgassing is in a narrow
sunward jet with a higher outgassing speed (0.9~\kms) than outside the jet 
(0.5~\kms).

\begin{acknowledgements} 
   
HIFI has been designed and built by a consortium of institutes and university
departments from across Europe, Canada and the United States under the
leadership of SRON Netherlands Institute for Space Research, Groningen, The
Netherlands and with major contributions from Germany, France and the US.
Consortium members are: Canada: CSA, U. Waterloo; France: CESR, LAB, LERMA, IRAM;
Germany: KOSMA, MPIfR, MPS; Ireland: NUI Maynooth; Italy: ASI, IFSI-INAF,
Osservatorio Astrofisico di Arcetri-INAF; Netherlands: SRON, TUD; Poland: CAMK,
CBK; Spain: Observatorio Astron\'{o}mico Nacional (IGN), Centro de
Astrobiolog\'{\i}a (CSIC-INTA); Sweden: Chalmers University of Technology --
MC2, RSS \& GARD; Onsala Space Observatory; Swedish National Space Board,
Stockholm University -- Stockholm Observatory; Switzerland: ETH Zurich, FHNW;
USA: Caltech, JPL, NHSC. We are grateful to the IRAM staff and to other
observers for their assistance during the observations. IRAM is an international
institute co-funded by the Centre national de la recherche scientifique (CNRS),
the Max Planck Gesellschaft and the Instituto Geogr\'afico Nacional, Spain. This
research has been supported by the CNRS and the Programme national de
plan\'etologie de l'Institut des sciences de l'univers (INSU). The CSO is
supported by National Science Foundation grant AST-0540882. The James Clerk
Maxwell Telescope is operated by The Joint Astronomy Centre on behalf of the
Science and Technology Facilities Council of the United Kingdom, the Netherlands
Organisation for Scientific Research, and the National Research Council of
Canada. The research leading to these results received funding from the European
Community's Seventh Framework Programme (FP7/2007--2013) under grant
agreement No. 229517. S.S. was supported by MNiSW (grant 181/N-HSO/2008/0).
\end{acknowledgements}

\end{document}